\documentstyle[12pt]{article}
\parskip=\medskipamount

\def\appendix#1{
\addtocounter{section}{1}
\setcounter{equation}{0}
\renewcommand{\thesection}{\Alph{section}}
\section*{Appendix \thesection\protect\indent #1}
\addcontentsline{toc}{section}{Appendix \thesection\ \ \ #1}
}
\newcommand{\tr}[1]{\,{\rm tr}\,#1\,}

\def\e{\varepsilon}

\def\be{\begin{equation}}
\def\la{\label}
\def\ee{\end{equation}}
\def\bea{\begin{eqnarray}}
\def\eea{\end{eqnarray}}
\def\eps{\varepsilon}
\def\a{\alpha}
\def\b{\beta}
\def\s{\sigma}
\def\ss{{\bf s}}
\def\n{\nabla}

\def\D{\Delta}

\def\g{\gamma}
\def\d{\delta}

\def\l{\left(}
\def\ll{\lambda}
\def\r{\right)}
\def\p{\partial}
\def\x{\vec{x}}
\def\y{\vec{y}}
\def\z{\vec{z}}
\def\vk{\vec{k}}

\newcommand{\cL}{{\cal L }}
\newcommand{\N}{{\cal N }}
\begin{document}
\begin{titlepage}
\thispagestyle{empty}
\title{
\large{
\begin{flushright}
LMU-TPW 00-8\\
UAHEP 00-3\\
hep-th/0003038
\end{flushright} }
\vskip 0.5cm
{\bf \huge{ 
On the correspondence between gravity fields
and CFT operators  
}}}  
\author{G. Arutyunov$^{a\, c}$
\thanks{arut@theorie.physik.uni-muenchen.de} \mbox{}
 and \mbox{} S. Frolov$^{b\,c}$
\thanks{frolov@bama.ua.edu 
\newline
$~~~~~$$^c$On leave of absence from 
Steklov Mathematical Institute, Gubkin str.8, GSP-1, 
117966, Moscow, Russia
}
\vspace{0.4cm} \mbox{} \\
\small {$^a$ Sektion Physik,}
\vspace{-0.1cm} \mbox{} \\
\small {Universit\"at M\"unchen,}
\small {Theresienstr. 37,}
\vspace{-0.1cm} \mbox{} \\
\small {D-80333 M\"unchen, Germany}
\vspace{0.4cm} \mbox{} \\
\small {$^b$ Department of Physics and Astronomy,}
\vspace{-0.1cm} \mbox{} \\
\small {University of Alabama, Box 870324,}
\vspace{-0.1cm} \mbox{} \\
\small {Tuscaloosa, Alabama 35487-0324, USA}
\mbox{}
}
\date {}
\maketitle
\begin{abstract}
It is shown that a nonlinear derivative-dependent transformation of gravity
fields changes correlation functions in a boundary CFT, and, 
therefore, corresponds to a change of a basis of operators 
in the CFT. It is argued that only non-renormalized
structures in correlation functions can be changed by such a
field transformation, and that the study of the response of correlation
functions to gravity field transformations allows one to find
them. In the case of 4-point
functions of CPOs in SYM$_4$ several non-renormalized structures 
are found, including the extremal and subextremal ones.
It is also checked that quartic 
couplings of scalar fields $s^I$ that are dual to extended chiral 
primary operators vanish in the subextremal case, as dictated by 
the non-renormalization theorem for the subextremal 4-point functions 
and the AdS/CFT correspondence.
\end{abstract}
\end{titlepage}
\newpage
\section{Introduction}
According to the AdS/CFT correspondence \cite{M,GKP,W} 
fields of type IIB supergravity on the $AdS_5\times S^5$ 
background are dual to gauge invariant operators in  
$D=4$, ${\cal N}=4$ supersymmetric Yang-Mills theory (SYM$_4$)
which belong to short representations of the conformal
superalgebra $SU(2,2|4)$ and have protected scale dimensions.
The short representations are generated by chiral primary
operators (CPOs)  transforming in the $k$-traceless symmetric
representations of $SO(6)$.
It is well-known that
single-trace operators 
$O^I_k = \tr (\phi^{(i_1}\cdots\phi^{i_k)})$ are chiral, and it was 
shown in \cite{AFer} that multi-trace operators of the form
$O^I_{k_1\cdots k_n} = \tr (\phi^{(i_1}\cdots\phi^{i_{k_1}})\cdots
\tr (\phi^{j_{1}}\cdots\phi^{j_{k_n})})$ are chiral too. 
There are also CPOs which are normal-ordered products 
of single- and 
multi-trace CPOs and their descendents.
Thus, in general, CPOs are admixtures of single- and multi-trace
operators with the same (protected) conformal dimension.

On the other hand the particle spectrum of type IIB supergravity on
$AdS_5\times S^5$ \cite{KRN,GM} contains only one set of fields 
which can couple to CPOs. These fields $s^I$ are mixtures of the
five form field strength and the trace of the graviton on the sphere.
Thus, one should understand which linear combinations of 
CPOs are dual to the gravity fields $s^I$.
Although, it is customarily believed that they are dual to single-trace 
operators $O^I_k$, no complete reliable proof of this fact is known.

A way to solve the problem is to compute correlation functions 
in free field theory and in the supergravity approximation, and to compare
them. Of course, one can compare only correlation functions 
subject to non-renormalization theorems. According to \cite{GKP,W}, to
compute n-point functions in SYM$_4$ one has to know the type 
IIB supergravity action on $AdS_5\times S^5$ up to the $n$-th order.
The quadratic action for physical fields was found in \cite{AF3} by
using the ``covariant'' action of \cite{ALS,ALT}.
The first step in finding interaction vertices 
was made in \cite{LMRS} where
quadratic and cubic actions for the scalars $s^I$ were found by 
expanding the covariant equations of motion \cite{S,SW,HW} 
for type IIB supergravity
up to the second order. By using
the actions, all 3-point functions of normalized CPOs dual to $s^I$ 
were computed, and, for generic values of conformal dimensions of 
the CPOs, appeared to coincide with 3-point functions of the 
single-trace CPOs $O^I_k$ calculated in the free field theory.
It was conjectured in \cite{LMRS} that the 3-point functions 
are not renormalized, and this was later proven in \cite{EHW}.
One might conclude on the basis of this coincidence that the 
fields $s^I$ are dual to the single-trace CPOs. However, as was
noted in \cite{HF1}, a 3-point function of CPOs computed 
in the supergravity approximation vanishes in the extremal case, 
for which the sum of conformal dimensions of two operators equals 
the conformal dimension of the third operator, e.g. $k_1=k_2+k_3$,
because of the vanishing of the cubic couplings of the dual
scalar fields.

There were proposed three different ways to resolve the puzzle. 
According to \cite{LT2}, to compute extremal 3-point
functions, one should first analytically continue in the
conformal dimensions $k_1,k_2,k_3$. Then, since the gravity
coupling is proportional to $k_2+k_3-k_1$, and the AdS integral 
\cite{FMMR} behaves itself as $1/(k_2+k_3-k_1)$, one obtains a finite
extremal 3-point function. However, from the computational point of
view the procedure of analytical continuation
looks superfluous, because no actual singularity
is involved. An extremal 3-point function
vanishes due to the absence of the corresponding cubic coupling, and 
one does not have to evaluate any AdS integral. 

In \cite{AF5} we explained  
the vanishing of the extremal cubic 
couplings by noting that the scalars $s^I$,
and, in general, supergravity fields, may be dual to extended CPOs
which are admixtures of single- and multi-trace CPOs. 
Nevertheless, the fact that the analytical continuation procedure 
seems to work in all known examples,\footnote{In particular, 
this procedure works in the case of 3-point functions of
operators dual to two scalars $s^I$ and a supergravity field, computed in
\cite{AF5}, where the cubic couplings \cite{AF5, Lee} also vanish
in extremal cases.} allows one to assume that, in the large $N$ limit
and for generic values of conformal dimensions, correlation functions
of extended CPOs coincide with the ones of the single-trace CPOs.
However, it is also clear that the analytical continuation procedure 
may work only in the large $N$ limit, because for finite $N$ 
only the single-trace CPOs $O^I_k$ with $2\le k\le N$ are independent, 
and a single-trace CPO with $k>N$ is equal to a linear combination
of multi-trace CPOs. This also shows that the appearence of
multi-trace CPOs is unavoidable for finite $N$.

Other arguments in favour of the proposal come from the study
of quartic couplings of the scalars $s^I$ performed in \cite{AF6}.
It is shown there that the quartic couplings
vanish in the extremal case for which 
$k_1=k_2+k_3+k_4$. As was pointed out in \cite{AF5} 
the vanishing of extremal couplings is dictated by the AdS/CFT 
correspondence because in this case
contact Feynman diagrams are ill-defined, and 
therefore, non-vanishing extremal quartic couplings 
would contradict to the AdS/CFT correspondence. 
By the same reason 2- and 4-derivative quartic couplings have to vanish
in the subextremal case for which $k_1=k_2+k_3+k_4-2$, and
4-derivative quartic couplings should vanish in the sub-subextremal
case when $k_1=k_2+k_3+k_4-4$.
The vanishing of extremal couplings means that 
4-point extremal correlators
of CPOs dual to the scalars $s^I$ vanish, and, therefore,
the scalars correspond not to single-trace CPOs but to extended CPOs. 

Then, it is shown in \cite{AF6} that the quartic action is consistent
with the Kaluza-Klein (KK) reduction down to five dimensions, and
admits a truncation to the massless multiplet, 
which can be identified with the field 
content of the gauged ${\cal N}=8$, $d=5$ supergravity \cite{PPN,GRW}.
Consistency means that there is no term linear in 
massive KK modes in the untruncated supergravity action, 
so that all massive KK fields can be put to zero without any
contradiction with equations of motion.
From the AdS/CFT correspondence point of view the consistent 
truncation implies that $any$ $n$-point correlation function 
of $n-1$ operators dual to the fields from the massless multiplet
and one operator dual to a massive KK field vanishes because, as one can 
easily see there is no exchange Feynman diagram in this case.   
This in particular implies that the scalars $s^I$ 
are dual to extended CPOs.
Indeed, if we assume that the scalars $s^I$ correspond to the single-trace
CPOs $O^I_k$,  we derive from the consistency
of the KK reduction that correlators of the form
$\langle O_2^{I_1}O_2^{I_2}\cdots O_2^{I_{n-1}}O_k^{I_n}\rangle$ 
vanish for $k\ge 4$, that is not the case for
such correlators of single-trace CPOs. 

Finally, the third way of solving the puzzle was proposed in \cite{DFMMR},
where it was noted that since the scalars $s^I$ used in \cite{LMRS}
differ from the original scalars appearing in the covariant equations
of motion for type IIB supergravity, 
one could obtain a nonvanishing extremal 3-point function by using
an action for the original scalars. This was demonstrated  for 
fields from the descendent sector where the relevant part of the type IIB
supergravity action is known. The action used in \cite{DFMMR} contains
higher-derivative terms, and although the bulk extremal couplings vanish
on shell, there appear boundary terms which provide nonvanishing
contribution to extremal 3-point functions.
However,  as was also noted in \cite{DFMMR}, one can make a nonlinear
off-shell transformation of the gravity fields and remove all
higher-derivative terms and all nonvanishing (off-shell) extremal
couplings. No boundary terms appear 
as a result of the field transformation,
and the transformed action leads to vanishing extremal correlators.

Thus, these results seem to indicate that although the
original gravity fields may be dual to single-trace operators, the
transformed fields are already dual to mixtures of single- and 
multi-trace operators. From this point of view a redefinition 
of the gravity fields corresponds to a change of an operator 
basis in CFT. 

To justify this point of view we study how 4-point correlation functions 
are changed under derivative-dependent gravity fields redefinitions 
of the form used in \cite{AF6} to reduce the non-Lagrangian equations 
of motion to a Lagrangian form. The field transformations 
discussed in \cite{DFMMR} are their particular case.
We begin with the quartic action for scalars $s^I$ found in \cite{AF6} and 
show that these transformations indeed can
change some correlators, in particular, the extremal 3- and 4-point 
functions and the subextremal 4-point functions for which
$k_1=k_2+k_3+k_4-2$. 

The subextremal correlators are of special interest
because, as has been shown in \cite{EHSSW2}, and checked 
in \cite{EP} to first order in perturbation theory,
they are not renormalized. 
Thanks to the non-renormalization theorem one can also employ
subextremal 4-point correlators to test the AdS/CFT correspondence. 
In particular the non-renormalization 
implies that the subextremal 
quartic couplings vanish,
and we show that this is indeed the case. 
This fact together with the absence of the exchange
Feynman diagrams leads to the vanishing of the 
subextremal 4-point functions of extended CPOs dual to the scalars $s^I$. 

We show that any field redefinition induces a change of correlation 
functions which is always
given by a product of 2- and 3-point functions. By this reason, and due to
the non-renormalization theorems for extremal and subextremal
correlators,
it seems possible to find such a field transformation that
the extremal and subextremal 3- and 4-point functions coincide 
with the ones of single-trace CPOs. 

As a by-product of our study, 
we also find that if conformal dimensions of at least two 
CPOs do not coincide then some structures in a 4-point function
of these CPOs can be also changed by a field
redefinition. Thus, although such a 4-point function in general is 
not protected by a non-renormalization theorem,
this  seems to be an indication that the coefficients of
the changing structures of the 
4-point function are not renormalized. 

The plan of the paper is as follows. In section 2 we recall the 
definition of normalized single-trace CPOs and extended CPOs, and discuss
the general properties of the supergravity action used to compute 3- 
and 4-point functions of the extended CPOs. In section 3 we study 
how the 3- and 4-point correlation functions change under a 
derivative-dependent field redefinition.
In appendix we show that the quartic couplings of \cite{AF6} 
vanish in the subextremal case, and that 4-derivative couplings vanish
in the sub-subextremal case.
\section{Extended CPOs and quartic supergravity action}
\setcounter{equation}{0}
We follow \cite{LMRS} defining the normalized single-trace CPOs as 
\bea
O^I(\x )=\frac{(2\pi )^k}{\sqrt{k\lambda^k}}C^I_{i_1\cdots i_k}
\tr (\phi^{i_1}(\x )\cdots \phi^{i_k}(\x )),
\la{cpo}
\eea
where $C^I_{i_1\cdots i_k}$ are totally symmetric traceless rank $k$ 
orthonormal tensors of $SO(6)$: 
$\langle C^IC^J\rangle =C^I_{i_1\cdots i_k}C^J_{i_1\cdots i_k}=\delta^{IJ}$, 
and $\phi^{i}$ are scalars of SYM$_4$. 

The two- and three-point functions of CPOs can be easily 
computed in free field theory and in the large $N$ limit, and  
are given by \cite{LMRS}
\bea
&&\langle O^I(\x )O^J(\y )\rangle =\frac{\delta^{IJ}}{|\x -\y |^{2k}},
\la{cpo2}\\
&&\langle O^{I_1}(\x )O^{I_2}(\y )O^{I_3}(\z )\rangle =\frac 1N
\frac{C^{I_1I_2I_3}}
{|\x -\y |^{2\a_3}|\y -\z |^{2\a_1}|\z -\x |^{2\a_2}},
\la{cpo3}
\eea
where $\a_i =\frac 12 (k_j+k_l-k_i)$, $j\neq l\neq i$, 
$C^{I_1I_2I_3}= 
\sqrt{k_1k_2k_3}\langle C^{I_1}C^{I_2}C^{I_3}\rangle $,
and 
$\langle
C^{I_1}C^{I_2}C^{I_3}\rangle $ is the unique $SO(6)$ invariant obtained 
by contracting $\a_1$ indices between $C^{I_2}$ and  $C^{I_3}$, 
$\a_2$ indices between $C^{I_3}$ and  $C^{I_1}$, and
$\a_3$ indices between $C^{I_2}$ and  $C^{I_1}$.
As was discussed in the Introduction, single-trace CPO 
cannot be dual to the scalar fields $s^I$ used in \cite{LMRS} to compute
their 3-point functions. However, as was shown in
\cite{AF5}, one can define
an extended CPO which corresponds to a scalar  $s^{I}$ by adding to 
a single-trace CPO a proper combination of multi-trace CPOs:
\bea
\tilde {O}^{I_1}=O^{I_1} -\frac {1}{2N}\sum_{I_2+I_3=I_1}
C^{I_1I_2I_3}O^{I_2}O^{I_3}.
\la{ecpo}
\eea
One can easily check that in the large $N$ limit these
operators have the normalized two-point functions (\ref{cpo2}),
the three-point functions (\ref{cpo3}) in the non-extremal case,
and vanishing three-point functions in the extremal case.  
Note that these operators require further modification to be 
consistent with all $n$-point functions computed in the framework of 
the AdS/CFT correspondence. 
In general, an extended CPO is a linear combination of a 
CPO and chiral primary composite operators which are normal-ordered 
products of CPOs and their descendants.
 
The quartic action for the scalars $s^I$ dual to 
the extended CPOs was found
in \cite{AF6}, and the part of the action
depending only on the scalars can be written in the form
\bea
S&=&\int_{AdS_5}\,\biggl( -\frac12 (\n_as_I\n^as_I +m_I^2s_I^2) + 
 \ll_{IJK} s_Is_Js_K +\ll_{IJKL}^{(0)} s_Is_Js_Ks_L \cr
&+&  
\ll_{IJKL}^{(2)} \n_as_I\n^as_Js_Ks_L +\ll_{IJKL}^{(4)}
 \n_as_I\n^as_J\n_bs_K\n^bs_L \biggr) \cr
&=& \int_{AdS_5}\,\cL (s^I).
\la{l1}
\eea
Since the action does not contain higher-derivative terms, 
the Hamiltonian reformulation of the quartic action
is straightforward, and, therefore, as was shown in \cite{AF1}, there is 
no need to add boundary terms. 

There are also cubic terms describing the interaction of the scalars $s^I$ 
with other scalars, with vector fields, and with massive symmetric
tensor fields of the second rank, but we omit them for the sake of 
simplicity.

Considering the contribution of contact 
Feynman diagrams to 3- and 4-point
functions, one can easily observe that 
the integrals over the $AdS_5$ space
diverge in several cases: 
$(i)$ if cubic couplings do not vanish in
the extremal case for which, e.g. $k_1=k_2+k_3$,
$(ii)$ if quartic couplings do 
not vanish in the extremal case when  $k_1=k_2+k_3+k_4$, 
$(iii)$ if 4-derivative and 2-derivative 
quartic couplings do not vanish in
the subextremal case when $k_1=k_2+k_3+k_4-2$, and $(iv)$ if 
4-derivative quartic couplings do not vanish in the sub-subextremal 
case for which $k_1=k_2+k_3+k_4-4$. 
Thus the AdS/CFT
correspondence requires vanishing all these couplings. 
Moreover, 
although the AdS integral involved in the subextremal non-derivative
quartic graph does not diverge, the non-derivative quartic couplings
also have to vanish in the subextremal case, because as was proven in 
\cite{EHSSW2} the subextremal 4-point functions are non-renormalized, 
and, therefore, have a free field form
 (a product of 2- and 3-point functions of a free CFT). 
On the other hand a nonvanishing
quartic subextremal coupling would lead to a 4-point function which 
does not have a free field form, and, this would contradict to the
AdS/CFT correspondence. 

Since one can easily show that all exchange Feynman diagrams
vanish in the extremal and subextremal cases, the vanishing of the
quartic couplings means that extremal and subextremal 4-point functions
of operators dual to the scalars $s^I$ in (\ref{l1})
also vanish. This is certanly not the case for the correlators 
of the single-trace CPOs $O^I_k$,
and, therefore, we interpret the scalars $s^I$
as to be dual to the extended CPOs of the form (\ref{ecpo}).
However, to obtain action (\ref{l1}) a number of nonlinear 
derivative-dependent field redefinitions was performed.
Thus, a natural question arises whether it is possible 
to make such a field redefinition of the 
scalars $s^I$ that the redefined scalars $\ss^I$ would correspond
to the single-trace CPOs. In the next section
we study the response of 3- and 4-point correlation functions to 
such changes and show that the desirable field redefinitions  
may exist.
\section{Field redefinitions and 4-point functions} 
\setcounter{equation}{0}
According to the proposal by \cite{GKP,W}, the generating functional
of connected Green functions in SYM$_4$ at large $N$ and at strong
't Hooft coupling coincides with the on-shell value of the type
IIB supergravity action on $AdS_5\times S^5$ subject to the Dirichlet
boundary conditions imposed on supergravity fields at the 
boundary of $AdS_5\times S^5$. To have a well-defined functional
of the boundary fields we cut the AdS space\footnote{
We use the AdS metric of the form: $ds^2=\frac{1}{z^2}(dz^2+dx_i^2)$.} 
 off at $z=\e$
and consider the part of AdS with $z\ge\e$. We impose the Dirichlet
boundary conditions on the scalars $s^I$: $s^I(\e ,\x )\equiv s^I(\x )$,
and denote the on-shell value of the action (\ref{l1}) as 
$S(s)$. To compute 3- and 4-point functions   
we only need equations of motion for the scalars $s_I$ decomposed 
up to the second order in fields:
\be
(\n_a^2 -m_I^2)s_I + 3\ll_{IJK}s_Js_K =0.
\la{eq1}
\ee
The solution of the equation that satisfies the Dirichlet boundary 
conditions can be written in the form
\be
s_I=s_I^{(0)}+ s_I^{(1)} .
\la{si}
\ee
Here $s_I^{(0)}$ solves the linear part of (\ref{eq1}) with 
the Dirichlet boundary conditions at $z=\e$, and $s_I^{(1)}$ 
has the vanishing boundary conditions  at $z=\e$, and solves the equation
\be
(\n_a^2 -m_I^2)s_I^{(1)} + 3\ll_{IJK}s_J^{(0)}s_K^{(0)} =0.
\la{eq11}
\ee
The on-shell value of action (\ref{l1}) is obtained 
by substituting (\ref{si}) into it:
\be
S(s)=\int_{\e}^\infty dz\,\int d\x\, z^{-5}\cL (s_I^{(0)}+s_I^{(1)}).
\la{s1}
\ee
Let us now consider the following off-shell transformation of the fields $s_I$
\bea
s_I&=&\ss_I+C_{IJK}^{(0)}\ss_J\ss_K+C_{IJK}^{(2)}\n_a\ss_J\n^a\ss_K
\cr
&+&C_{IJKL}^{(0)}\ss_J\ss_K\ss_L+C_{IJKL}^{(2)}\n_a\ss_J\n^a\ss_K\ss_L
+C_{IJKL}^{(4)}\n_a\ss_J\n_b\ss_K\n^a\n^b\ss_L\cr
&=&\ss_I+(\d s)_I.
\la{redefin}
\eea
This transformation is of the same form as the most general 
$s$-dependent field redefinition that was used in \cite{AF6} to reduce
the original non-Lagrangian equations of motion to a Lagrangian form.
We also assume that
the constants $C_{IJK}^{(0)}$ and $C_{IJK}^{(2)}$ do not vanish only
if any of the conformal dimensions does not exceed its extremal value:
$\D_I\le\D_J+\D_K$.

The equations of motion for the redefined fields look as follows
\be
(\n_a^2 -m_I^2)\ss_I + 3\ll_{IJK}\ss_J\ss_K 
+(\n_a^2 -m_I^2)(\d_2 s)_I=0,
\la{eq2}
\ee
where
\be
(\d_2 s)_I=C_{IJK}^{(0)}\ss_J\ss_K+C_{IJK}^{(2)}\n_a\ss_J\n^a\ss_K .
\la{d2s}
\ee
The simplest way to study the influence of the field 
redefinition (\ref{redefin}) on the 3- and 4-point functions is to
impose on the redefined fields $\ss_I$ the same boundary conditions
as on $s_I$, and to compare the on-shell values of the
original and transformed actions. 
Thus we write the solution to (\ref{eq2}) in the form
\be
\ss_I=s_I^{(0)}+ \ss_I^{(1)}
=s_I^{(0)}+ s_I^{(1)} - (\d_2 s)_I^{(0)}+\s_I^{(0)}.
\la{ssi}
\ee
Here
\be
(\d_2 s)_I^{(0)}=C_{IJK}^{(0)}s_J^{(0)}s_K^{(0)}
+C_{IJK}^{(2)}\n_as_J^{(0)}\n^as_K^{(0)},
\ee
and $\s_I^{(0)}$ solves the linear part of the equation  (\ref{eq1}), 
and satisfies the following boundary condition
\be
\s_I^{(0)}|_{z=\e}=(\d_2 s)_I^{(0)}|_{z=\e}.
\la{boun}
\ee
Substituting (\ref{ssi}) into (\ref{redefin}), we find
\bea
\ss_I+(\d s)_I=s_I^{(0)}+ s_I^{(1)} +\s_I^{(0)} +(\d_3 s)_I^{(0)}+
2C_{IJK}^{(0)}s_J^{(0)}\ss_K^{(1)}
+2C_{IJK}^{(2)}\n_as_J^{(0)}\n^a\ss_K^{(1)},
\nonumber
\eea
where
\bea
(\d_3 s)_I^{(0)}
=C_{IJKL}^{(0)}s_J^{(0)}s_K^{(0)}s_L^{(0)}
+C_{IJKL}^{(2)}\n_as_J^{(0)}\n^as_K^{(0)}s_L^{(0)}
+C_{IJKL}^{(4)}\n_as_J^{(0)}\n_bs_K^{(0)}\n^a\n^bs_L^{(0)}.
\nonumber
\eea
Thus the on-shell value of the transformed Lagrangian is given by
\bea
\tilde{\cL}(\ss_I)&=&\cL (\ss_I +(\d s)_I)=\cL (s_I^{(0)}+ s_I^{(1)})\cr
&-&\n_a \left( s_I^{(0)}+ s_I^{(1)}\right)\n^a \left(\s_I^{(0)} 
+(\d_3 s)_I^{(0)}+
2C_{IJK}^{(0)}s_J^{(0)}\ss_K^{(1)}
+2C_{IJK}^{(2)}\n_bs_J^{(0)}\n^b\ss_K^{(1)}\right)\cr
&-&m_I^2\left( s_I^{(0)}+ s_I^{(1)}\right)\left(\s_I^{(0)} +(\d_3 s)_I^{(0)}+
2C_{IJK}^{(0)}s_J^{(0)}\ss_K^{(1)}
+2C_{IJK}^{(2)}\n_as_J^{(0)}\n^a\ss_K^{(1)}\right)\cr
&-&\frac12 \n_a\s_I^{(0)}\n^a\s_I^{(0)}-
\frac12 m_I^2\s_I^{(0)}\s_I^{(0)}+3\ll_{IJK}s_I^{(0)}s_J^{(0)}\s_K^{(0)}.
\la{onsl}
\eea
The first term on the r.h.s.  of this equation is equal to the on-shell 
value of the original Lagrangian (\ref{l1}), and other terms represent 
relevant corrections to it. By using equations of motion we can rewrite 
(\ref{onsl}) as follows
\bea
\tilde{\cL}(\ss_I)&=&\cL (s_I^{(0)}+ s_I^{(1)})\cr
&-&\n_a\biggl(\n^a  \left(
s_I^{(0)}+s_I^{(1)}\right)
\left(\s_I^{(0)}+(\d_3 s)_I^{(0)}+
2C_{IJK}^{(0)}s_J^{(0)}\ss_K^{(1)}
+2C_{IJK}^{(2)}\n_bs_J^{(0)}\n^b\ss_K^{(1)}\right)\biggr)\cr
&-&
\frac12 \n_a\biggl(\n^a \s_I^{(0)}\s_I^{(0)}\biggr).
\la{ansl1}
\eea
Thus the on-shell values of the original action and the redefined one only
differ by a boundary term. Omitting nonessential terms, which cannot
change the 3- and 4-point functions,
this boundary term can be written in the form
\bea
I=\int\,d\x\, \e^{-d+1} &\biggl(& \e^2C_{IJK}^{(2)}\p_0s_I^{(0)}
\p_0s_J^{(0)}\p_0s_K^{(0)}
+  \e^2C_{IJK}^{(2)}\p_0s_I^{(1)}\p_0s_J^{(0)}\p_0s_K^{(0)}\cr
&+& \p_0s_I^{(0)}\left( C_{IJKL}^{(2)}\n_as_J^{(0)}
\n^as_K^{(0)}s_L^{(0)} + C_{IJKL}^{(4)}\n_as_J^{(0)}
\n^bs_K^{(0)}\n^a\n_bs_L^{(0)}\right)\cr
&+&2\e^2C_{IJK}^{(2)}\p_0s_I^{(0)}
\p_0s_J^{(0)}\p_0\left( s_K^{(1)}-(\d_2 s)_K^{(0)}+\s_K^{(0)}\right)+
\frac12 \p_0 \s_I^{(0)}\s_I^{(0)}\biggr) .
\la{bterm}
\eea
The first term on the r.h.s. of (\ref{bterm}) represents 
the change in a 3-point function
induced by the field redefinition (\ref{redefin}). It was shown  
in \cite{DFMMR} that this term gives a nonvanishing contribution only to
the extremal 3-point functions, and always leads to a contribution 
of the free-field form.
In particular, one can choose a field redefinition of such a form
that all 3-point functions will coincide with the 3-point functions of
the single-trace CPOs.

We are going to study the influence of the field redefinition on 
the 4-point functions and 
begin with considering the simplest term\footnote{For the sake of generality
we consider the boundary term (\ref{bterm}) in $d$ dimensions, and
for arbitrary scalars $s_I$ dual to operators of conformal dimensions
$\D_I$ with the only restriction $\D_I\ge d/2$. }
\be
I_1=\int\, d\x \,\e^{-d+1}  C_{IJKL}^{(2)}\p_0s_I^{(0)}
\n_as_J^{(0)}\n^as_K^{(0)}s_L^{(0)}.
\la{I1}
\ee
It is obvious that only derivatives in the radial direction can give a
nonvanishing contribution to a 4-point function, thus we can replace 
$I_1$ by
\be
I_1=\int\, d\x \, \e^{-d+3}  C_{IJKL}^{(2)}\p_0s_I^{(0)}
\p_0s_J^{(0)}\p_0s_K^{(0)}s_L^{(0)}.
\la{I11}
\ee
It is well-known that the Fourier transform of 
the solution of the Dirichlet problem is given by (see, e.g. \cite{DFMMR})
\bea
s_I^{(0)}(z,\vk )=K_I(z,\vk )s_I(\vk ),
\nonumber
\eea
where
\be
K_I(z,\vk )=\l\frac{z}{\e}\r^{d/2}
\frac{{\cal K}_\nu (kz)}{{\cal K}_\nu (k\e )},~~\nu =\D_I -\frac{d}{2}
\ee
and ${\cal K}_\nu (kz)$ is a Macdonald function.
By using this formula, we find
\be
z\p_0 K_I(z,\vk )|_{z=\e}=d-\D_I 
+ a_I (k\e )^{2(\D_I-\frac{d}{2})}\log (k\e ) +\cdots ,
\la{exp}
\ee
where $\cdots$ refer to terms which do not contribute to 4-point functions
in the limit $\e\to 0$.
Thus a relevant contribution of $I_1$ to the Fourier transform of a 
4-point function is proportional
to 
\bea
 \d (\vk_1 +\vk_2 +\vk_3 +\vk_4 )\e^{2\D_I+2\D_J+2\D_K-4d}
k_1^{\D_I -\frac{d}{2}}k_2^{\D_J -\frac{d}{2}}k_3^{\D_K -\frac{d}{2}}
\log (k_1\e )\log (k_2\e )\log (k_3\e ).
\nonumber
\eea
This expression is always the Fourier transform of 
a product of three 2-point functions. It gives a 
contribution to a 4-point function only if \footnote{Note that the correct
scaling behaviour of a $4$-point function is 
${\cal O}\l\e^{\D_I+\D_J+\D_K+\D_L -4d}\r$.}
$$
2\D_I+2\D_J+2\D_K -4d=\D_I+\D_J+\D_K+\D_L -4d,
$$
i.e., in the extremal case $\D_I+\D_J+\D_K=\D_L$.
Note that in the non-extremal cases 
$\D_L < \D_I+\D_J+\D_K $, in particular in the subextremal one, 
the boundary term scales too fast to give a contribution. 

The second integral to be considered is
 \be
I_2=
\int\,d\x\,  \e^{-d+3}  C_{IJL}^{(2)}\p_0s_I^{(0)}
\p_0s_J^{(0)}\p_0\s_L^{(0)}.
\la{I2}
\ee
To analyse the integral we use
\bea
\s_L^{(0)}(z,\vk )&=&K_L(z,\vk )\int\, d\vk_3d\vk_4\,\d (\vk_3+\vk_4-\vk)
\biggl( 
C_{LMN}^{(2)}\e^2\p_0 s_M^{(0)}(\e ,\vk_3 )\p_0s_N^{(0)}(\e ,\vk_4 ) \cr
&+&
C_{LMN}^{(2)}\e^2 \p_is_M(\vk_3 )\p_is_N(\vk_4 ) +
C_{LMN}^{(0)}s_M(\vk_3 )s_N(\vk_4 )\biggr) .
\la{sig}
\eea
One can easily see that only the first term can give a nonvanishing 
contribution to a 4-point function. Combining (\ref{I2}) with (\ref{sig})
we get that this contribution to a 4-point function is proportional to
\bea
&&C_{IJL}^{(2)} C_{LMN}^{(2)}\d (\vk_1+\vk_2+\vk_3+\vk_4 ) \e^{-d+5} 
\p_0K_I(\e ,\vk_1 )\p_0K_J(\e ,\vk_2 )\p_0K_L(\e ,\vk_3+ \vk_4)\cr
&&\times  \p_0 K_M(\e ,\vk_3 )\p_0K_N(\e ,\vk_4 ).
\la{I22}
\eea
By using (\ref{exp}), we see that there are several cases when we can get a 
nonlocal contribution: $(i)$ the five-logs case, $(ii)$ the four-logs case,
and $(iii)$ the three-logs case. It is not difficult to show that there is 
no contribution in the five- and four-logs cases.  
Three-logs case has three subcases. The first one is
$$
\log (k_3+k_4)\log k_3\log k_4\d (\vk_1+\vk_2+\vk_3 +\vk_4 )
$$
In this case we get $\d (x_2-x_1)$ after integrating over momenta.

The second case is
$$
\log k_2\log k_3\log k_4\d (\vk_1+\vk_2+\vk_3 +\vk_4 )
$$
It is obvious that in this case we get a product of three 2-point 
functions, and a nonvanishing contribution will be only 
in the extremal case $\D_J+\D_M+\D_N =\D_I$.

The third case is
$$
\log (k_3+k_4)\log k_2\log k_4\d (\vk_1+\vk_2+\vk_3 +\vk_4 )
$$
One can easily see that in this case we also obtain
a product of three 2-point 
functions, and a nonvanishing contribution will be only if
$$
-\D_I+\D_J+2\D_L-\D_M+\D_N =0.
$$
Taking into account that
$$
-\D_I+\D_J+\D_L\ge 0,\qquad -\D_M+\D_N+\D_L\ge 0
$$
we get that there is a solution to this equation if
$$
\D_L=\D_I-\D_J=\D_M-\D_N .
$$
This is a new case, and it is tempting to assume that the 
coefficient of the 
corresponding structure in the 4-point function is not renormalized.
To understand why this may be so it is instructive 
to write down the changing term in the 4-point function:
\bea
\langle O^{I}(\x_1)O^{J}(\x_2)O^{M}(\x_3)O^{N}(\x_4)\rangle 
\sim\frac{1}{x_{12}^{2\D_J}x_{13}^{2\D_L}x_{34}^{2\D_N}}+\cdots .
\nonumber
\eea
We see that this term is obtained by plunging the operators $O^J$ and $O^N$
into the operators $O^I$ and $O^M$ respectively. The perturbative
non-renormalization of the Feynman diagrams of such a type 
was checked in \cite{EP} to first order in perturbation theory
where it was noted that this effectively is 
equivalent to the proof of the non-renormalization of 2-point functions
of CPOs given in \cite{DFS}.

The next integral to be considered is
\be
I_3=
\int\,d\x \,  \e^{-d+3}  C_{IJL}^{(2)}\p_0s_I^{(0)}
\p_0s_J^{(0)}\p_0s_L^{(1)} .
\la{I3}
\ee
Taking into account that $s_L^{(1)}$ solves the equation (\ref{eq11}),
we obtain the formula
$$
s_L^{(1)}(x_0,\x )=3\lambda_{LMN}\int\, d^{d+1}y\, \sqrt{g}G_L^{\e}(x,y)
 s_M^{(0)}(y)s_N^{(0)}(y) ,
$$
where the Green function can be found in \cite{MV1}, and
satisfies
\bea
\frac{\p }{\p x_0}G_L^{\e}(x,y)|_{x_0=\e }
=-\e^{d-1}
\int\frac{d\vk}{(2\pi )^d}e^{-i\vk (\x -\y )}
K_L(y_0,\vk )=-\eps^{\D_L-1}K_{\D_L}(y_0,\x-\y),
\nonumber
\eea
where $K_{\D_L}(x_0,\x-\y)$ is the bulk-to-boundary 
propagator defined in \cite{FMMR}.

It is convenient to analyse (\ref{I3}) in the x-space, where
the solution $s_I^{(0)}$ can be written as 
\bea
s_I^{(0)}(x_0,\x)=\frac{1}{\eps^{d-\D_I}}\l \int d\y K_{\D_I}(x_0,\x-\y)s(\y)
+o(\eps ) \r . 
\eea

Thus, for $s_L^{(1)}$ we have 
\bea
&&\p_0s_L^{(1)}(\eps,\x)=-3\lambda_{LMN} \eps^{\D_L+\D_M+\D_N-2d-1}\times\\
&&
\int d\y_3 d\y_4 s_M(\y_3)s_N(\y_4)\l \int d^{d+1}y 
\sqrt{g}K_{\D_L}(y_0,\x-\y) K_{\D_M}(y_0,\y-\y_3)
K_{\D_N}(y_0,\y-\y_4)+o(\eps )\r .
\nonumber
\eea
Since the cubic couplings $\ll_{LMN}$ vanish in the extremal case, 
the integral 
$$
\int_{\eps}^{\infty}dy_0 \int d\y \sqrt{g}K_{\D_L}(y_0,\x-\y)
K_{\D_M}(y_0,\y-\y_3)K_{\D_N}(y_0,\y-\y_4),
$$
which appears in evaluation of a 3-point function,
is finite in the limit $\eps\to 0$ (it diverges only in the extremal case) 
and, therefore, can be approximated as 
$$
\Lambda_{LMN}(\x;\y_3,\y_4)+o( \eps ),
$$
where $\Lambda_{LMN}(\x,\y_3,\y_4)$ is defined as 
$$
\Lambda_{LMN}(\x;\y_3,\y_4)
=\int_{0}^{\infty}dy_0 \int d\y  \sqrt{g}K_{\D_L}(y_0,\x-\y) 
K_{\D_M}(y_0,\y-\y_3)K_{\D_N}(y_0,\y-\y_4). 
$$
Thus, $\p_0s_L^{(1)}$ behaves itself as 
\bea
\p_0s_L^{(1)}(\eps,\x)=-3\lambda_{LMN}\eps^{\D_L+\D_M+\D_N-2d-1}
\l A_{LMN}(\x)+ o(\eps) \r 
\eea
with $A_{LMN}(\x)=\int d\y_3d\y_4 s(\y_3)s(\y_4)\Lambda_{LMN}(\x;\y_3,\y_4)$ 
and we, therefore, find 
$$
I_3=-3C_{IJL}^{(2)}\lambda_{LMN}\eps^{\D_L+\D_M+\D_N-3d}\int d\x~~
\eps\p_0s_I^{(0)}\eps\p_0s_J^{(0)}(A_{LMN}(\x)+o(\eps)\ ).
$$
The last formula allows one to determine 
the behavior of the corresponding correlation 
function in the momentum space. Namely, the leading in $\eps$
contribution to the 4-point correlation function is proportional to 
\bea
\la{lc}
C_{IJL}^{(2)}\lambda_{LMN}\eps^{\D_L+\D_M+\D_N-3d}~~
\eps\p_0 K_I(\eps,\vk_1)\eps\p_0 K_J(\eps,\vk_2)
\Lambda_{LMN}(\vk_1+\vk_2;\vk_3,\vk_4),  
\eea
where function $\Lambda_{LMN}(\vk ; \vk_3,\vk_4)$ 
stands now for the Fourier transform of 
$\Lambda_{LMN}(\x;\y_3,\y_4)$ in all its arguments.

To find the relevant contribution to the 4-point function 
we use (\ref{exp}) so that  
the leading contribution (\ref{lc}) is given by 
the sum of two different terms:
the first one contains only one log, while the second one  
contains the product of two logs. We first consider the one-log
case and show that it provides in particular 
a contribution to a subextremal 
4-point correlation function.
For definiteness we pick up here the following term 
\bea
\la{lcsub}
C_{IJL}^{(2)}\lambda_{LMN}\eps^{\D_L+\D_M+\D_N+2\D_J-4d}
k_2^{2\D_J-d}\log k_2\Lambda_{LMN}(\vk_1+\vk_2;\vk_3,\vk_4),
\eea       
which gives a nonvanishing contribution if
$$
2\D_J+\D_L+\D_M+\D_N-4d=\D_I+\D_J+\D_M+\D_N-4d
$$
i.e. $\D_I=\D_J+\D_L$. Clearly, this equality is not too restrictive and 
it allows in particular the solution for the subextremal 
case\footnote{The extremal case is of no interest here 
since the coupling $\lambda_{LMN}$ vanishes.}, 
i.e., when $\D_I=\D_J+\D_M+\D_N-2$ and $\D_L=\D_M+\D_N-2$.
Let us now show that for these values of conformal dimensions 
(\ref{lcsub}) indeed represents the relevant momentum space structure
of a subextremal 4-point correlation function.

Due to the non-renormalization theorem a subextremal 
4-point correlation function of single-trace CPOs is given by the sum 
of products of two-point functions 
that is further restricted by the conformal invariance to the form
\bea
\langle O^{I}(\x_1)O^{J}(\x_2)O^{M}(\x_3)O^{N}(\x_4)\rangle 
=\frac{A_{IJMN}}{x_{12}^{2(\D_J+\a-1)}x_{13}^{2(\D_M+\b-1)}
x_{14}^{2(\D_N+\g-1)}
x_{23}^{2\g}x_{24}^{2\b}x_{34}^{2\a}},
\eea
where $\D_I=\D_J+\D_M+\D_N-2$ 
and $\a,\b,\g$ are integers obeying the condition $\a+\b+\g=1$,
so that only one of them is non-zero and equals to 1. Thus we have 
three different subextremal structures which have in general three 
different coefficients $A$.  
 
Consider the structure with $\b=\g=0$ and perform the Fourier transform.
The corresponding structure in the momentum space looks as 
\bea
\nonumber
&&\langle O^{I}(\vk_1)O^{J}(\vk_2)O^{M}(\vk_3)O^{N}(\vk_4)\rangle \sim
\int d\x_1d\x_2d\x_3d\x_4 
\frac{e^{i\vk_1\x_1+i\vk_2\x_2+i\vk_3\x_3+i\vk_4\x_4} }
{x_{12}^{2\D_J}x_{13}^{2(\D_M-1)}x_{14}^{2(\D_N-1)}
x_{34}^{2} } \\
\nonumber
&& \sim \d(\vk_1+\vk_2+\vk_3+\vk_4)k_2^{2\D_J-d}\log k_2 
\int dv dw\frac{e^{-i\vk_3 v- i\vk_4 w}}{v^{2(\D_M-1)}w^{2(\D_N-1)}(v-w)^2 },
\eea 
where new integration variables $v=x_{13}$ and $w=x_{14}$ were introduced.
Coming back to (\ref{lcsub}) it remains to note that in 
the x-space  $\Lambda_{LMN}(\x;\y_3,\y_4)$
is fixed by conformal invariance to be 
\bea
\nonumber
\Lambda_{LMN}(\x;\y_3,\y_4)=\frac{C}{(\x-\y_3)^{2(\D_M-1)}
(\x-\y_4)^{2(\D_N-1)}(\y_3-\y_4)^2},
\eea
where we have used subextremality condition $\D_L=\D_M+\D_N-2$ and $C$
is the numerical (non-zero) constant. Transforming this expression 
to the momentum space we therefore find 
\bea
\nonumber
\Lambda_{LMN}(\vk_1+\vk_2;\vk_3,\vk_4)=\d(\vk_1+\vk_2+\vk_3+\vk_4)
\int dv dw\frac{e^{-i\vk_3 v- i\vk_4 w}}{v^{2(\D_M-1)}w^{2(\D_N-1)}(v-w)^2}.
\eea 
Thus, we have shown that the field redefinition induces 
a non-trivial contribution to the subextremal 4-point functions.

The general case  $\D_I=\D_J+\D_L$ is considered in the same way, and
we get that the changing term in a 4-point function has the form
\bea
\langle O^{I}(\x_1)O^{J}(\x_2)O^{M}(\x_3)O^{N}(\x_4)\rangle 
\sim\frac{1}{x_{12}^{2\D_J}x_{13}^{\D_L +\D_M -\D_N}
x_{14}^{\D_L +\D_N -\D_M}x_{34}^{\D_M +\D_N -\D_L}
} +\cdots .
\nonumber
\eea
We see that this term is obtained by plunging the operator $O^J$
into the operator $O^I$. The perturbative
non-renormalization of the Feynman diagrams of such a type 
seems to be equivalent to the proof of the non-renormalization of 
2- and 3-point functions of CPOs given in \cite{DFS}. 

Consideration of the term involving two logs shows that it scales too fast 
and by this reason does not lead to any contribution to 4-point functions.

The last integral to be considered is
\be
I_4=\int\, d\x \,\e^{-d+1}  C_{IJKL}^{(4)}\p_0s_I^{(0)}
\n_as_J^{(0)}\n^bs_K^{(0)}\n^a\n_bs_L^{(0)}
\la{I4}
\ee
The only case when the integral gives a contribution to a 4-point function
is $a=b=0$. However, in this case we can use the equations of motion
for scalars $s^I$ to express $\n^0\n_0s_L^{(0)}$ as 
$(m_L^2-\n^i\n_i )s_L^{(0)}$. Thus, this integral is equivalent to
the integral $I_1$ (\ref{I1}), and can give a nonvanishing 
contribution to an extremal 4-point function.

This completes our consideration of the boundary terms (\ref{bterm}).
\section{Conclusion}
In this paper we studied nonlinear derivative-dependent transformations
of gravity fields, and showed that they change 
3- and 4-point functions in a boundary CFT. 
We interpreted such a change of correlation functions as 
a manifestation of an operator basis transformation in CFT.
Thus, a derivative-dependent field redefinition invokes
a transformation of operators in CFT, and, as the consequence, 
a transformation of correlation functions. 
However, this transformation has a very restrictive form 
as by a derivative-dependent field redefinition
it is possible to change only 
the coefficients of non-renormalized structures 
of correlation functions.
In particular, one probably can find such a field redefinition
of the gravity fields that the redefined scalars $s^I$ would be dual
to the single-trace CPOs. Still the analysis performed in the
paper does not allow one to conclude this definitely. The point is
that we do not have enough parameters in the 
field transformations because  
we only considered field redefinitions of the scalars $s^I$.
In general, one should take into account the scalar dependent 
redefinitions of vector and tensor fields as well. 
This would give us enough parameters to 
transform the correlation functions of the extended CPOs into
the ones of the single-trace CPOs.  
\section{Appendix}
\setcounter{equation}{0}
In \cite{AF6} the quartic action for scalars $s^I$ was found in the form 
\bea
\nonumber
S(s)=\int_{AdS_5}\left(\cL_4^{(4)}+\cL_4^{(2)}+ \cL_4^{(0)}\right),
\eea 
where the quartic terms contain the 4-derivative couplings
\bea
\nonumber
\cL_4^{(4)}=
\l S_{I_1I_2I_3I_4}^{(4)}
+A_{I_1I_2I_3I_4}^{(4)}\r s^{I_1}\n_as^{I_2}\n_b^2(s^{I_3}\n^as^{I_4}),
\eea
the 2-derivative couplings 
\bea
\cL_4^{(2)}=
\l S_{I_1I_2I_3I_4}^{(2)}
+A_{I_1I_2I_3I_4}^{(2)}\r s^{I_1}\n_as^{I_2}s^{I_3}\n^as^{I_4}
\nonumber
\eea
and the couplings without derivatives 
\bea
&&\cL_4^{(0)}=S_{I_1I_2I_3I_4}^{(0)}~s^{I_1}s^{I_2}s^{I_3}s^{I_4}.
\nonumber
\eea
The corresponding vertices have the following symmetry properties 
\bea
\nonumber
&&S_{I_1I_2I_3I_4}^{(4)}=S_{I_2I_1I_3I_4}^{(4)}=S_{I_3I_4I_1I_2}^{(4)},~~~
A_{I_1I_2I_3I_4}^{(4)}=-A_{I_2I_1I_3I_4}^{(4)}=A_{I_3I_4I_1I_2}^{(4)}, \\
\nonumber
&&S_{I_1I_2I_3I_4}^{(2)}=S_{I_2I_1I_3I_4}^{(2)}=S_{I_3I_4I_1I_2}^{(2)},~~~
A_{I_1I_2I_3I_4}^{(2)}=-A_{I_2I_1I_3I_4}^{(2)}=A_{I_3I_4I_1I_2}^{(2)}
\eea
and their explicit values are given in \cite{AF6}. What is important 
for our discussion here is that all the couplings are  
represented as sums of the $SO(6)$ tensors of three different types:
\bea
\nonumber
F(I_5)a_{I_1I_2I_5}a_{I_3I_4I_5},~~~
F(I_5)t_{I_1I_2I_5}t_{I_3I_4I_5},~~~
F(I_5)p_{I_1I_2I_5}p_{I_3I_4I_5},
\eea  
where $F(I_5)$ is a function of $I_5$ and the sum over $I_5$ is assumed. There 
also appear tensors obtained from these ones by different permutation 
of indices. Recall that $a_{I_1I_2I_3}$,  $t_{I_1I_2I_3}$
and $p_{I_1I_2I_3}$ represent the following integrals involving 
the scalar $Y^{I}$, the vector $Y_\a^{I}$ and the tensor $Y_{(\a\b )}^{I}$
spherical harmonics respectively
\bea
\nonumber
a_{I_1I_2I_3}=\int_{S^5}~ Y^{I_1}Y^{I_2}Y^{I_3}, ~~~~
t_{I_1I_2I_3}=\int_{S^5}~ \n^\a Y^{I_1}Y^{I_2}Y_\a^{I_3},~~~~
p_{I_1I_2I_3}=\int_{S^5}~ \n^\a Y^{I_1}\n^\b Y^{I_2}Y_{(\a\b )}^{I_3}.
\eea

To prove the vanishing of the couplings in the subextremal case 
as well as the vanishing of the 4-derivative couplings in the sub-subextremal 
case we find convenient to pass to the 4-derivative vertices  
of the form (\ref{l1}). 
This is achieved by using the following relations valid on-shell:
\bea
\nonumber
A^{(4)}_{1234}\int s_1\n_a s_2 \n_b^2(s_3\n^a s_4)&=&
-2A^{(4)}_{1234}\int \n_a s_1\n_b s_2 \n^a s_3\n^b s_4 \\
\nonumber
&-&4A^{(4)}_{1234} \int s_1\n_a s_2 s_3\n^a s_4\\\nonumber
&-&\frac{1}{4}A^{(4)}_{1234}(m_1^2-m_2^2)(m_3^2-m_4^2) 
\int s_1 s_2 s_3 s_4.\\\nonumber
S^{(4)}_{1234}\int s_1\n_a s_2 \n_b^2(s_3\n^a s_4)&=&
-S^{(4)}_{1234}\int \n_a s_1\n^a s_2 \n_b s_3\n^b s_4 \\
\nonumber
&+&S^{(4)}_{1234}\l m_1^2+m_2^2+m_3^2+m_4^2-4 \r
\int s_1\n_a s_2 s_3\n^a s_4 \\
\la{1der}
&+&\frac{1}{4}S^{(4)}_{1234}(m_1^2+m_2^2)(m_3^2+m_4^2) 
\int s_1 s_2 s_3 s_4,
\eea
where here and below we write concisely the summation index $I_1$ simply as 1
and similar for the others, $m$ denotes the $AdS$ mass of a scalar field.  

First we consider the subextremal case and assume for definiteness that 
$k_1=k_2+k_3+k_4-2$. It is easy to
show, by using the description of spherical harmonics as restrictions
of functions, vectors and tensors on the ${\bf R}^6$ in which the sphere $S^5$ 
is embedded \cite{LMRS,AF5}, that the tensor\footnote{We do not assume here 
summation over $I_5$.} $t_{125}t_{345}$ does not vanish in the subextremal 
case only for $k_5=k_3+k_4-1$ while for $p_{125}p_{345}$ it is the case 
only if $k_5=k_3+k_4-2$. As for the tensor $a_{125}a_{345}$,
it differs from zero in two cases: when $k_5=k_3+k_4$ or
when $k_5=k_3+k_4-2$. Analogously, the only non-trivial values of $k_5$
for $a_{135}a_{245}$ are $k_5=k_2+k_4$ and $k_5=k_2+k_4-2$,
and for $a_{145}a_{235}$ they are $k_5=k_2+k_3$ and $k_5=k_2+k_3-2$.
Thus in all vertices we can replace $k_5$ by a corresponding 
function of $k_2,k_3,k_4$, and,
then the only dependence on $k_5$ is in tensors $t_{125}t_{345}$,
$p_{125}p_{345}$, $a_{125}a_{345}$, $a_{135}a_{245}$ and $a_{145}a_{235}$.
However, not all of these tensors are independent. Indeed,
$a_{125}a_{345}$, $a_{135}a_{245}$ and $a_{145}a_{235}$ are 
subjected to the following three identities \cite{AF6}: 
\bea
\la{ida}
&&a_{125}a_{345}=a_{135}a_{245}=a_{145}a_{235},\\
\nonumber
&& f_5(a_{125}a_{345}+a_{135}a_{245}+a_{235}a_{145})=(f_1+f_2+f_3+f_4)
a_{125}a_{345},
\eea 
where $f_i=f(k_i)=k_i(k_i+4)$. 
For the sake of simplicity it is useful to introduce 
the notation  
\bea
\nonumber
\begin{array}{ll}
l_1=a_{125}a_{345}|_{k_5=k_3+k_4},&~ l_2=a_{125}a_{345}|_{k_5=k_3+k_4-2}, \\
m_1=a_{145}a_{235}|_{k_5=k_2+k_3},&~ m_2=a_{145}a_{235}|_{k_5=k_2+k_3-2}, \\
n_1=a_{135}a_{245}|_{k_5=k_2+k_4},&~ n_2=a_{135}a_{245}|_{k_5=k_2+k_4-2},
\end{array}
\eea
where, e.g., $l_1$ denotes tensor $a_{125}a_{345}$ for the value of $k_5$
equal to $k_3+k_4$. Hence we have six tensors corresponding to different 
values of $k_5$ and to different order of indices, which are confined 
by three relations (\ref{ida}). Therefore, restricting eqs.(\ref{ida})
to the subextremal case, i.e., putting $k_1=k_2+k_3+k_4-2$ 
one can solve them for any three tensors. If we choose
here $l_1$, $m_1$ and $n_1$ as independent variables, then  
$l_2$, $m_2$ and $n_2$ are expressed as
\bea
\nonumber
l_2&=&\frac{(m_1 + n_1 - l_1)(k_2+1) + m_1 k_3 + n_1 k_4}
{k_1 + k_2 + k_3+2},\\
\la{subrel}
m_2&=&\frac{(n_1 + l_1 -m_1)(k_4+1)+ l_1k_3 + n_1k_2}{k_1 + k_2 + k_3+2},\\
\nonumber
n_2&=&\frac{(m_1 - n_1 + l_1)(k_3+1) + l_1k_4 + m_1 k_2}{k_1 + k_2 + k_3+2}.
\eea  

For $t_{125}t_{345}$ and $p_{125}p_{345}$ we will need the following three
identities found in \cite{AF6}:
\bea
&&t_{125}t_{345}=-\frac{(f_1-f_2)(f_3-f_4)}{4f_5}a_{125}a_{345}
+\frac{1}{4}f_5(a_{145}a_{235}-a_{245}a_{135}),
\la{TT}\\
&&(1-f_5)t_{125}t_{345}=\frac{1}{4}(f_5^2-f_5(f_1+f_2+f_3+f_4-4))
(a_{145}a_{235}-a_{135}a_{245})
\la{TTF}\\
&&\qquad-\frac{4-f_5}{4f_5}(f_1-f_2)(f_3-f_4)a_{125}a_{345} , \nonumber\\
&&p_{125}p_{345}=
-\frac{(f_1-f_2)(f_3-f_4)}{2(f_5-5)}t_{125}t_{345}
-\frac{5}{4f_5(f_5-5)}d_{125}d_{345}\nonumber\\ 
&&\qquad-\frac{1}{20}(f_1+f_2-f_5)(f_3+f_4-f_5)a_{125}a_{345}
+\frac{1}{8}(f_1+f_3-f_5)(f_2+f_4-f_5)a_{135}a_{245}
\nonumber\\
\la{PP}
&&\qquad+\frac{1}{8}(f_1+f_4-f_5)(f_2+f_3-f_5)a_{145}a_{235},
\eea
where  
$$
d_{123}=\int_{S^5}\n^{(\a}\n^{\b)}Y^{I_3}\n_{\a}Y^{I_1}\n_{\b}Y^{I_2}=
\biggl( \frac{1}{10}f_2f_3+\frac{1}{10}f_1f_3+\frac{1}{2}f_1f_2-\frac{1}{4}f_1^2
-\frac{1}{4}f_2^2+\frac{3}{20}f_3^2\biggl)a_{125}.
$$
Since in the subextremal case $t_{125}t_{345}$ is non-zero only for one 
value of $k_5$ we may use formula (\ref{TT}) and eqs.(\ref{subrel})
to express $t_{125}t_{345}$ in terms of $l_1$, $m_1$ and $n_1$.
Similarly, combining eq.(\ref{PP}) with (\ref{TT}) and with eqs.(\ref{subrel})
one obtains an analogous representation for $p_{125}p_{345}$. 
In this way we have expressed all the quartic vertices via independent tensors
$l_1$, $m_1$ and $n_1$. 

Now we single out the field $s^{I_1}$ and write the relevant part
of the quartic 4-derivative vertices as functions of $l_1$, $m_1$ and $n_1$
in the form
\bea
\L^{(4)} = 4\sum_{I_2,I_3,I_4}\biggl( -S_{I_1I_2I_3I_4}^{(4)}
-A_{I_1I_3I_2I_4}^{(4)}+A_{I_1I_4I_3I_2}^{(4)}\biggr)
\n_a s^{I_1}\n^as^{I_2}\n_bs^{I_3}\n^bs^{I_4}, \la{4}
\eea
where we sum over the representations satisfying the subextremality
condition. Now, we substitute the values of $k_5$ discussed above, and
$k_1=k_2+k_3+k_4-2$ in the quartic couplings, and obtain zero.

To analyse 2-derivative terms we represent the 2-derivative Lagrangian
as follows
\bea
\L^{(2)} &=& 4\sum_{I_2,I_3,I_4}\biggl( 
\l -\frac12\tilde{S}_{I_1I_3I_2I_4}^{(2)}
+\tilde{A}_{I_1I_2I_3I_4}^{(2)}\r 
s^{I_1}\n^as^{I_2}s^{I_3}\n_as^{I_4} \nonumber\\
&+&\frac14\l \tilde{A}_{I_1I_2I_3I_4}^{(2)}(m_4^2-m_3^2)-
\tilde{S}_{I_1I_2I_3I_4}^{(2)}(m_4^2+m_3^2)\r 
s^{I_1}s^{I_2}s^{I_3}s^{I_4}\biggr) ,
\nonumber
\eea
where using (\ref{1der}) we define
\bea
\tilde{S}_{I_1I_2I_3I_4}^{(2)}&=&S_{I_1I_2I_3I_4}^{(2)}
+S^{(4)}_{I_1I_2I_3I_4}\l m_1^2+m_2^2+m_3^2+m_4^2-4 \r ,\nonumber\\
\tilde{A}_{I_1I_2I_3I_4}^{(2)}&=&A_{I_1I_2I_3I_4}^{(2)}
-4A^{(4)}_{I_1I_2I_3I_4}.
\nonumber
\eea
This time substituting $k_5$ and $k_1$ and symmetrizing the expression 
obtained in $I_2$ and $I_4$, we get a non-zero function which is, however,
completely symmetric in $I_2$, $I_3$ and $I_4$. Thus we can remove the
2-derivative term by using the shift
$$
s^{I_1}\to s^{I_1} -\frac{2}{3\kappa_1} \l 
-\frac12\tilde{S}_{I_1I_3I_2I_4}^{(2)}
+\tilde{A}_{I_1I_2I_3I_4}^{(2)}\r s^{I_2}s^{I_3}s^{I_4},$$
where $\kappa_1=\frac{32k_1(k_1-1)(k_1+2)}{k_1+1}$.
This shift also produces an additional contribution to the non-derivative 
terms which is equal to
$$ -\frac23 \l 
-\frac12\tilde{S}_{I_1I_3I_2I_4}^{(2)}
+\tilde{A}_{I_1I_2I_3I_4}^{(2)}\r (m_2^2+m_3^2+m_4^2-m_1^2)
s^{I_1}s^{I_2}s^{I_3}s^{I_4}.
$$

After accounting this contribution the non-derivative terms acquire the form
\bea
\L^{(0)} &=& 4\sum_{I_2,I_3,I_4}\biggl( S_{I_1I_2I_3I_4}^{(0)}-
\frac16
\l -\frac12\tilde{S}_{I_1I_3I_2I_4}^{(2)}
+\tilde{A}_{I_1I_2I_3I_4}^{(2)}\r (m_2^2+m_3^2+m_4^2-m_1^2) \nonumber\\
&+&\frac14\l \tilde{A}_{I_1I_2I_3I_4}^{(2)}(m_4^2-m_3^2)-
\tilde{S}_{I_1I_2I_3I_4}^{(2)}(m_4^2+m_3^2)\r \cr
&+&\frac{1}{4}S^{(4)}_{1234}(m_1^2+m_2^2)(m_3^2+m_4^2)
-\frac{1}{4}A^{(4)}_{1234}(m_1^2-m_2^2)(m_3^2-m_4^2) 
\biggr)
s^{I_1}s^{I_2}s^{I_3}s^{I_4}.
\nonumber
\eea
Substituting $k_5$ and $k_1$ and symmetrizing the coefficient 
obtained in $I_2$, $I_3$ and $I_4$ we end up with zero.
Thus, we have shown that after the additional field redefinition 
all subextremal quartic couplings vanish.

The treatment of the 4-derivative quartic couplings in the 
sub-subextremal case is quite analogous to the previous one.
For definiteness we assume that  $k_1=k_2+k_3+k_4-4$.
Then $a_{125}a_{345}$ is non-zero in three cases:
$k_5=k_3+k_4$, $k_5=k_3+k_4-2$ and $k_5=k_3+k_4-4$. Similarly 
$a_{135}a_{245}$ is non-zero only for $k_5$ equal to 
$k_2+k_4$, $k_2+k_4-2$ or $k_2+k_4-4$, while $a_{145}a_{235}$
admits for $k_5$ one of the following values 
$k_2+k_3$, $k_2+k_3-2$ or $k_2+k_3-4$. Denote 
\bea
\nonumber
\begin{array}{lll}
l_1=a_{125}a_{345}|_{k_5=k_3+k_4},&~ l_2=a_{125}a_{345}|_{k_5=k_3+k_4-2}, 
&~l_3=a_{125}a_{345}|_{k_5=k_3+k_4-4}, 
\\
m_1=a_{145}a_{235}|_{k_5=k_2+k_3},&~ m_2=a_{145}a_{235}|_{k_5=k_2+k_3-2}, 
&~m_3=a_{145}a_{235}|_{k_5=k_2+k_3-4}, 
\\
n_1=a_{135}a_{245}|_{k_5=k_2+k_4},&~ n_2=a_{135}a_{245}|_{k_5=k_2+k_4-2},
&~n_3=a_{135}a_{245}|_{k_5=k_2+k_4-4}.
\end{array}
\eea
Then identities (\ref{ida}) allow one to express $l_3,m_3,n_3$
via six independent tensors $l_1,m_1,n_1$ and $l_2,m_2,n_2$, e.g.,
\bea
l_3&=&-\frac{1}{k_2 + k_3 + k_4 +2}\biggl(
m_2 + n_2 + l_2+(- 2m_1 - m_2 + l_2 )k_3 \\
\nonumber
&+& (- 2m_1 - m_2 -2n_1
-n_2+ 2l_1+2l_2 )k_2+(- 2n_1- n_2+ l_2)k_4 \biggl).
\eea
The formulas for $m_3$ and $n_3$ are obtained from this one by 
permutations of indices. 

Except the tensor structures we have just considered the quartic
couplings of 4-derivative vertices  contain a tensor 
$(f_5-1)^2t_{125}t_{345}$ (see Appendix A of \cite{AF6}). 
In the sub-subextremal case 
$t_{125}t_{345}$ differs from zero only for $k_5=k_3+k_4-1$
or $k_5=k_3+k_4-3$. It is then suitable to represent 
\bea
\la{srep}
(f_5-1)^2t_{125}t_{345}=
\l (f_5-\a)(f_5-\b)+a(f_5-1)+b \r t_{125}t_{345} ,
\eea
where 
\bea
\nonumber
a&=&\a+\b-2 , \\
\nonumber
b&=&-(\a-1)(\b-1)
\eea
and pick up for $\a$ and $\b$ the following values  
$\a=f(k_3+k_4-1)$ and $\b=f(k_3+k_4-3)$. Clearly, in the sub-subextremal 
case the first term in the r.h.s. of (\ref{srep})
is absent and we may use identities (\ref{TT}) and (\ref{TTF})
to rewrite $(f_5-1)^2t_{125}t_{345}$ via $l_1,\ldots, n_2$. 
Hence, as in the subextremal case, we reduced all 
quartic couplings of 4-derivative vertices to the 
independent tensor structures.

Now upon substituting in eq.(\ref{4}) the 4-derivative quartic couplings 
evaluated for the proper values of $k_5$ and
putting $k_1=k_2+k_3+k_4-4$ we obtain zero. 

\vskip 1cm
{\bf ACKNOWLEDGMENT}

We would like to thank A. Tseytlin for prompt reading of 
the manuscript and useful remarks, and 
D. Freedman for useful correspondence,
G.A. is grateful to S. Theisen and S. Kuzenko, 
and S.F. is grateful to A. Tseytlin and S. Mathur
for valuable discussions. The work of G.A. was
supported by the Alexander von Humboldt Foundation and in part by the
RFBI grant N99-01-00166, and the work of S.F. was supported by
the U.S. Department of Energy under grant No. DE-FG02-96ER40967 and
in part by RFBI grant N99-01-00190.

\end{document}